\documentclass[aps,twocolumn,prb,10pt]{revtex4-1}

\usepackage{graphicx}

\begin{document}

\author{Cezary \'Sliwa}
\affiliation{Institute of Physics, Polish Academy of Sciences,
Aleja\ Lotnikow 32/46, PL-02668 Warsaw, Poland}

\title{Disorder-averaged Binder ratio in site-diluted Heisenberg models}
\date{2022-08-31}

\begin{abstract}
It is demonstrated via a numerical experiment (a Monte Carlo simulation) in the context of three-dimensional site-diluted Heisenberg spin systems
that a functional dependence of the Binder ratio ($V_4$) on the order parameter correlation length ($\xi / L$) requires a modification
to the usual definition of $V_4$ in disordered systems. An appropriate disorder averaging procedure is proposed.
\end{abstract}

\maketitle

\section{Introduction}

Physical systems investigated in statistical mechanics are frequently considered on a model level and treated with Monte Carlo approaches.
For example, magnets are modeled by systems of either Ising \cite{Onsager:1944} or Heisenberg \cite{Campos:2006,Lee:2007,Kawamura_2010,PhysRevB.102.174410} spins, interacting via distance-dependent exchange interactions.
Then, the critical properties one is interested in are usually described in terms of the so-called \emph{phenomenological couplings,} such as the Binder ratios of cumulants \cite{Binder:1981}.
However, when disorder is present --- here, site dilution --- a proper disorder average is a non-trivial and still open question.

A possible point of view --- the one followed by most researchers, is to interpret $V_4 = \frac{1}{2} \left\{ 3 - \left< m^4 \right> / \left( \left< m^2 \right> \right)^2 \right\}$ as
a simple transformation of the ratio $\left< m^4 \right> / \left( \left< m^2 \right> \right)^2$. In the latter expression, both numerator and denominator include
thermal averaged powers of the order parameter, $m$, in a combination which is system-size invariant. This leads to the following definition (for a single-component order parameter):
\begin{equation}
  V_4\{m\} = \frac{1}{2} \frac{3 \left[ \left< m^2 \right> \right]^2 - \left[ \left< m^4 \right> \right]}{\left[ \left< m^2 \right> \right]^2}.
  \label{eq: v4}
\end{equation}
($m$ is the order parameter, angle brackets $\left< \ldots \right>$ denote thermal expected value of an observable, and the square ones $\left[ \ldots \right]$ --- the average
over disorder realizations such as site-dilution configurations).
Alternatively, following the conviction that $V_4$ is a ratio of powers of magnetization cumulants,
and thus that of the free energy derivatives with respect to a probe magnetic field,
one performs disorder averages in the numerator and denominator as the last operation before the final division, as in:
\begin{equation}
  V_4'\{m\} = \frac{1}{2} \frac{\left[ 3 \left( \left< m^2 \right> \right)^2 - \left< m^4 \right> \right]}{\left[ \left( \left< m^2 \right> \right)^2 \right]}.
  \label{eq: v4'}
\end{equation}
\begin{figure}
\centerline{\includegraphics[width=0.98\columnwidth]{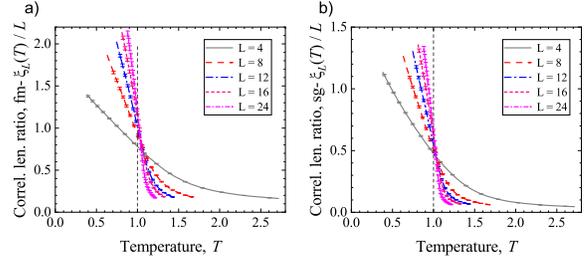}}
\caption{Correlation length ratios in the ferromagnetic and spin-glass sectors of a 3D site-diluted system of Heisenberg spins on the fcc lattice with $x = 0.3$, $J_1 = 1$, and $J_{2,3,4} = 0.1$.}
\label{fig: xi}
\end{figure}
\begin{figure}[b]
\centerline{\includegraphics[width=0.98\columnwidth]{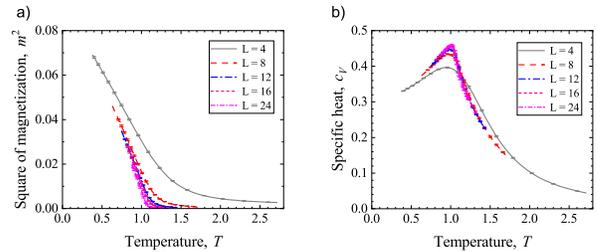}}
\caption{Magnetization $m$ (squared) and the specific heat $c_V$ (normalized by the number of lattice sites, $4 L^3$).}
\label{fig: cv}
\end{figure}
One therefore has the following relation to self-averaging, $R_{\chi}$
\begin{equation}
  \frac{3 - 2 V_4}{3 - 2 V_4'} = 1 + R_{\chi},
  \label{eq: reltosa}
\end{equation}
where (cf.\ Refs.\ \onlinecite{Aharony_1996,Aharony_1998,Gordillo_Guerrero_2007})
\begin{equation}
  R_{\chi} = \frac{\left[ \left( \left< m^2 \right> \right)^2 \right] - \left[ \left< m^2 \right> \right]^2}{\left[ \left< m^2 \right> \right]^2}.
\end{equation}
In order to eliminate the possibility of a divergence, it is probably preferrable to calculate $\tilde{V}_4\{m\}$ instead, defined as
\begin{equation}
  \tilde{V}_4\{m\} = \frac{\left[ \left< m^4 \right> - \left( \left< m^2 \right> \right)^2 \right]}{\left[ \left< m^4 \right> \right]}
\end{equation}
and directly related to $V_4'$. The definition implies a bound $0 \leq \tilde{V}_4\{m\} \leq 1$ (of course, the actual values in a particular system need not span the whole range from 0 to 1,
but they can be mapped to the $[0, 1]$ range by an appropriate linear transformation as in the case of $V_4$).
The purpose of the present work is to compare the two quantities, $V_4$ and $V_4'$, regarding their usefulness in Monte Carlo numerical experiments aiming
at the determination of the critical temperature in site-diluted Heisenberg spin systems.

\section{Correlation length on the lattice}

In this section we discuss how the discreteness of the lattice affects the calculation of the correlation length. For a simple cubic lattice (sc) it is customary to consider $\mathbf{k}_{min} = (0, 0, 1)$
(in units of $2 \pi / L$) as the minimum momentum in order to define
\begin{equation}
  \xi_{sc} = \frac{1}{2 \sin(k_{min} / 2)} \sqrt{\frac{\chi(\mathbf{k} = 0)}{\chi(\mathbf{k}_{min})} - 1}.
\end{equation}
We reconsider this definition emphasizing that the real space is a lattice with periodic boundary conditions on the $L \times L \times L$-block boundary. Let us assume that the spin-spin correlation function
assumes the ideal (but anisotropic) form:
\begin{equation}
  C(x, y, z) = e^{-(|x| + |y| + |z|) / \xi}.
  \label{eq: C x y z}
\end{equation}
Our goal is to recover the ``true'' value of $\xi$. We first sum up $C$ over the images, $C_L = \sum_{n_x, n_y, n_z} C(x + n_x L, y + n_y L, z + n_z L)$, then calculate the discrete Fourier transform $\tilde{C}_L$ according to $f(x) \mapsto \tilde f(k) = \sum_x f(x) e^{2 \pi i k x / L}$ (in each space direction). If $\tilde{C}_L(\mathbf{k})$ is substituted in place of $\chi(\mathbf{k})$, we obtain
\begin{equation}
  \xi_{sc} = \frac{1}{2 \sinh(1 / 2 \xi)},
\end{equation}
which implies a correction at small correlation lengths.

\begin{figure}
\centerline{\includegraphics[width=0.98\columnwidth]{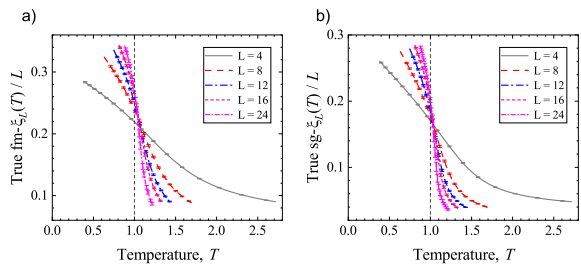}}
\caption{Corrected (``true'') correlation length ratios in the ferromagnetic and spin-glass sectors, cf.\ Fig.\ \ref{fig: xi}.}
\label{fig: true_xi}
\end{figure}

While repeating the same steps for the fcc lattice we have the choice of either $\mathbf{k}_{min} = (1, 1, 1)$ or $\mathbf{k}_{min}' = (0, 0, 2)$, then (respectively)
\begin{equation}
  \xi_{fcc} = \frac{1}{2 \sin(\pi / L)} \sqrt{\frac{\chi(\mathbf{k} = 0)}{\chi(\mathbf{k}_{min})} - 1}
\end{equation}
or
\begin{equation}
  \xi_{fcc}' = \frac{1}{4 \sin(\pi / L)} \sqrt{\frac{\chi(\mathbf{k} = 0)}{\chi(\mathbf{k}_{min}')} - 1}.
\end{equation}
Unexpectedly, the properties of the two definitions differ rather significantly, as $\xi_{fcc}$ is asymptotically rather like $\xi_{fcc} \propto \xi^3 / L^2$. To recover $\xi$, we solve the equation:
\begin{widetext}
\begin{eqnarray}
  \lefteqn{\Biggl[ \cos \frac{4\pi}{L} \left(7+\cosh\frac{1}{\xi}\right) - 2 \cos \frac{2\pi}{L} \left(7+\cosh\frac{1}{\xi}\right)\left(-1+3\cosh\frac{1}{\xi}\right) + {} \Biggr.} \nonumber \\
  & & \Biggl. {} + 3 \left(8 - 4\cosh\frac{1}{\xi}+3\cosh\frac{2}{\xi}+\cosh\frac{3}{\xi}\right) \Biggr] \left[ 32 \left(\sinh\frac{1}{2\xi}\right)^{6} \left(4+3\cos\frac{2\pi}{L}+\cosh\frac{1}{\xi}\right) \right]^{-1} = \xi_{fcc}^2
\end{eqnarray}
\end{widetext}
(the analogous equation involving $\xi_{fcc}'$ is somewhat simpler; on the other hand, $k_{min} < k_{min}'$).

\begin{figure*}
\centerline{\includegraphics[width=0.98\textwidth]{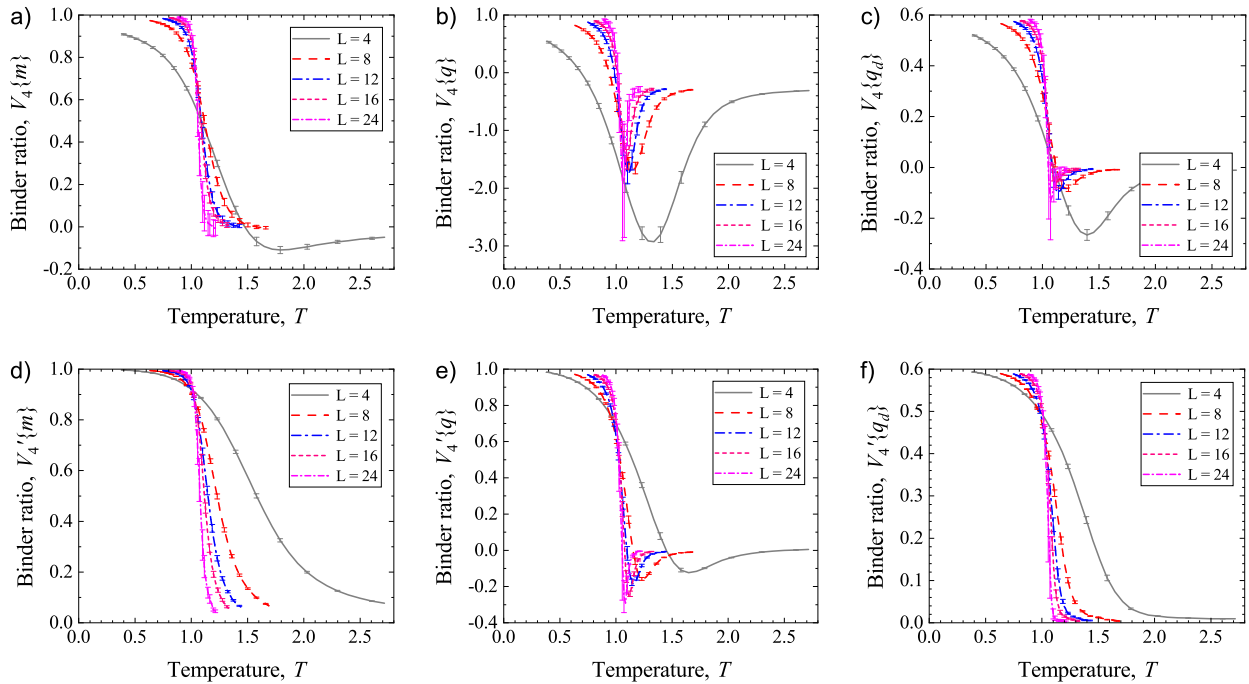}}
\caption{Binder ratios vs. temperature in the model system. Left column: ferromagnetic sector; middle column: spin-glass sector; right column: unitary spin-glass sector (see text).
Upper row: standard averaging ($V_4$), lower row: current definition ($V_4'$).}
\label{fig: v4}
\end{figure*}

\section{The model system and simulations}

We consider as our model a system of Heisenberg $O(n = 3)$ spins on a block of the thee-dimensional ($d = 3$) fcc lattice, with periodic boundary conditions.
The exchange couplings include ferromagnetic ones to the 12 nearest neighbors ($J_1 = 1$), and weaker couplings to the spins in the second, third, and fourth coordination spheres
($J_2 = 0.1$, $J_3 = 0.1$, $J_4 = 0.1$). The system is diluted, meaning that each lattice site is occupied with a probability $x$ (here, $x = 0.3$).
Such a model reflects the microscopic spin interactions in diluted magnetic semiconductors, where the signs of the exchange interactions (positive, $J > 0$, in a ferromagnet \cite{Stefanowicz_2013};
negative, $J < 0$, in an antiferromagnet or spin-glass \cite{Binder:1986}) are determined by the electronic configuration of the magnetic dopants constituting the spins.

In order to reduce computational complexity, usually tremendous in frustrated systems (antiferromagnetically-coupled spins on an fcc lattice),
only ferromagnetic couplings are considered here. However, 4 independent simulations are run simultaneously in order to compute the spin-glass order parameter $q$,
and parallel tempering \cite{Hukushima_1996} is active in order to achieve enhanced thermalization (see Appendix \ref{app: ptg}).

The set of system sizes is $L = 4, 8, 12, 16, 24$ (then the number of possibly occupied lattice sites is $4 L^3$).
For each system size, a set of $N_r$ ($N_r = 2048, 256, 128, 32, 8$, respectively) disorder configurations is generated,
and parallel tempering Monte Carlo simulations are run independently for each realization, with sequential \cite{sequpd:2006} sweeps of heat bath as the local update (see Appendix \ref{app: heat bath} for details of the algorithm) and each sweep being followed
by $2 L$ overrelaxation sweeps. Temperatures replicas ($T = T_1, T_2, \ldots, T_{N_T}$) are spaced in geometric progression (this would be adequate for a system with temperature-independent heat capacity,
as in the case of frustrated antiferromagnetic interactions, $J_i < 0$), with $N_T = 48, 60, 72, 96, 24$.
The first temperature of the second half of each set is $T_i = 1 / 0.96 \approx 1.04, i = \frac{N_T}{2} + 1$.
For each disorder configurations, $4 + 4$ measurements cycles are run, each consisting of $N_{mc}$ ($N_{mc} = 1250, 2500, 3750, 5000, 7500$) Monte Carlo steps both for
the burn-in or decorrelation phase and the proper measurement. Bias-free estimation of Binder ratios can be achieved by a combination of the prescription from Appendix\ \ref{app: rat_avg}
with standard jacknife resampling. Pseudo-random numbers are generated by the \texttt{xoshiro256**} algorithm \cite{Blackman_2021}.

\section{Results}

Figure \ref{fig: xi} shows the $L$ and temperature ($T$) dependence of $\text{fm-}\xi(T) / L$ and $\text{sg-}\xi(T) / L$, which are the ratio of the magnetization ($m$) and spin-glass ($q$)
correlation lengths $\xi$ to the system size. These quantities have been used to determine the Curie and spin-glass freezing transition temperature up to now.
The curves in the figure meet near $T \approx 1.0$, indicating the possibility of a phase transition at this temperature.
The order parameter changes its value (Fig.\ \ref{fig: cv}a) and the specific heat (Fig.\ \ref{fig: cv}b) peaks near this point.
A small shift between the common crossings of the curves in the two sectors is visible,
and it may be disputable whether the values of the corresponding transition temperatures ($T_c$ and $T_f$, respectively) are equal or not.

The corrected correlation lengths are shown in Fig.\ \ref{fig: true_xi}, which should be confronted with Fig.\ \ref{fig: xi}.
Since $\xi \propto L \iff \xi_{fcc} \propto L$, the correction does not shift much the points of common crossings
on the temperature scale, but the overall magnitude of the correlation lengths is diminished and the shape of the curves changes.

A similar picture is evident from the behavior of the Binder ratio $V_4$ in the ferromagnetic sector (Fig. \ref{fig: v4}, upper panel, left).
However, in the spin-glass sector (upper panel, middle), the corresponding Binder ratio features a well-known dip of negative values \cite{Fernandez:2007},
and its scaling properties are inconclusive (a common intersection cannot be located). Since the dip is not visible if individual disorder realizations are examined,
it appears only due to the averaging procedure implied by the definition of $V_4$. In contrast, the alternative definition ($V_4'$) yields
a well-defined common crossing in the spin-glass sector, as demonstrated in the lower panel of the Figure. A small dip is still visible;
this issue is addressed in the following Section\ \ref{sec: decomp_q} by selecting a component of the spin-glass order parameter $q$.
The Binder ratio curves for the selected component are shown in the right-most panel of the Figure (this component will be denominated as \emph{unitary}).

\begin{figure}
\centerline{\includegraphics[width=0.98\columnwidth]{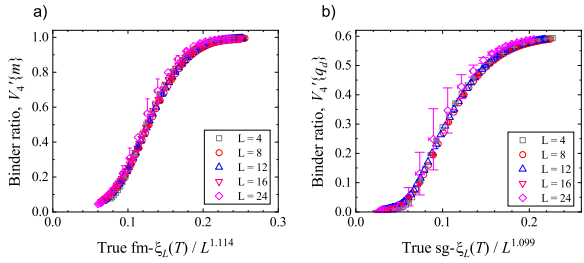}}
\caption{Binder ratios vs. the correlation length ratio in the model system. Left panel: ferromagnetic sector, right panel: unitary spin-glass sector.}
\label{fig: v4 vs xitol}
\end{figure}

\begin{figure}[b]
\centerline{\includegraphics[width=0.98\columnwidth]{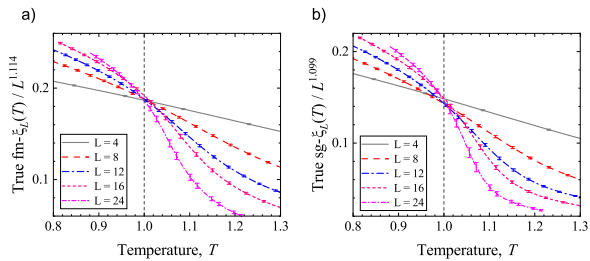}}
\caption{Determination of the critical (Curie) temperature from $\xi_L(T)$.}
\label{fig: crit}
\end{figure}

Finally, in order to verify that the modified (``alternative'') disorder average preserves the required properties of the Binder ratios,
one plots $V_4'$ versus the corresponding correlation length ratio (Fig. \ref{fig: v4 vs xitol}). An appropriate data collapse
onto one curve is visible only if the power of the linear system size in the ratio is slightly changed,
from $\xi_L / L^1$ to $\text{fm-}\xi_{s} = \xi_L / L^{1.114(2)}$ and $\text{sg-}\xi_{s} = \xi_L / L^{1.099(2)}$
(in the ferromagnetic and unitary spin-glass sectors, respectively), where the numerical values have been obtained by fitting a linear combination
of $1 / (1 + e^{-x})$ and $R_{2,2}(x) / \cosh(x / 2)$ [$x = a (\xi_L / L^y - \xi_{s,0})$ and $R_{2,2}$ is a rational function].
Corrections to scaling has thus become too small to be resolved at the present accuracy.
Now we can assume that $\xi_{s}$'s are the corresponding phenomenological couplings and take advantage of them in order to determine the critical temperature (Fig.\ \ref{fig: crit}).

\begin{figure}[b]
\centerline{\includegraphics[width=0.98\columnwidth]{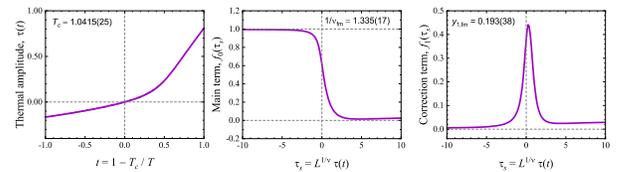}}
\caption{Amplitudes contributing to the fit of $V_4'(L, T)$ in the ferromagnetic sector (from left to right): thermal amplitude $\tau(t)$, main term $f_0(\tau_s)$, and the leading correction $f_1(\tau_s)$.}
\label{fig: fit fm}
\end{figure}

\begin{figure}
\centerline{\includegraphics[width=0.98\columnwidth]{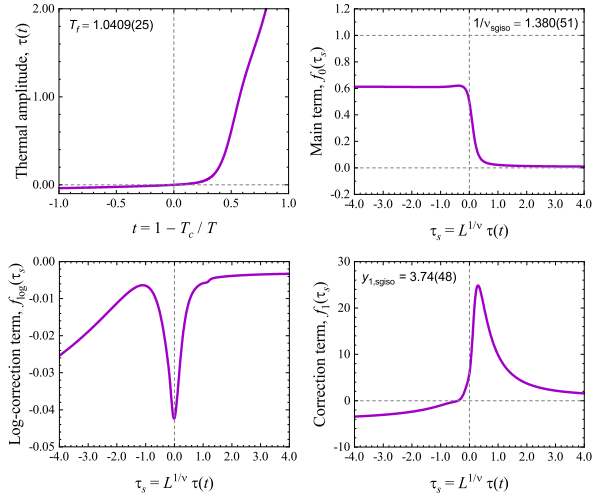}}
\caption{Amplitudes contributing to the fit of $V_4'(L, T)$ in the unitary spin-glass sector (in the reading order): thermal amplitude $\tau(t)$, main term $f_0(\tau_s)$, logarithmic-correction term $f_{\log}(\tau_s)$, and the first correction $f_1(\tau_s)$.}
\label{fig: fit sg}
\end{figure}

A more traditional (and quantitative approach) is to fit $V_4'(L, T)$ to the expression:
\begin{eqnarray}
  V_4'(L, T) & = & f_0(L^{1/\nu} \tau) + \log(L) f_{\log}(L^{1/\nu} \tau) + {} \nonumber \\
  & & \qquad {} + L^{-y_1} f_1(L^{1/\nu} \tau),
  \label{eq: fit1}
\end{eqnarray}
with arbitrary functions $f_0(\tau_s)$, $f_{\log}(\tau_s)$ (only in the spin-glass sector), $f_1(\tau_s)$, and $\tau(t)$,
\begin{equation}
 \tau = \tau(t \equiv 1 - T_c / T), \qquad L^{1/\nu} \tau \equiv \tau_s.
  \label{eq: fit2}
\end{equation}
In (\ref{eq: fit1}) and (\ref{eq: fit2}), $T_c$ is the critical temperature, $\nu$ is the thermal exponent,
and $y_1$ is the scaling exponent for the leading correction.
Converged nonlinear generalized least-squares fits have been obtained by running a piece of C++ code, linked to \texttt{mpfr} and implementing the BFGS algorithm (with a custom line-search routine). The fitted values satisfy \cite{Aharony_1996} the inequality $\nu > 2 / 3$ (alternative fits with $\nu < 2 / 3$ may also be found, but should be rejected).
Crout's algorithm with pivoting was used for matrix inversion. Second derivatives were calculated by combining symbolic expressions exported from Mathematica.
The values are $T_c = 1.0415(25)$, $1 / \nu_{\text{fm}} = 1.335(17)$, and $y_{1,\text{fm}} = 0.193(18)$ in the ferromagnetic sector (with $\chi^2 = 353.7$),
and $T_f = 1.0409(25)$, $1 / \nu_{\text{sgiso}} = 1.380(51)$, and $y_{1,\text{sgiso}} = 3.74(48)$ in the unitary spin-glass sector (with $\chi^2 = 293.7$).
The estimated covariance matrices for $L = 16$ (and $24$) were singular because of small $N_r$, and only every fourth (and sixth) temperature point of these curves could be included in the fit (then the number of degrees of freedom for the two fits, i.e.\ the numbers of data points less the number of parameters, are $208 - 34 = 174$ and $208 - 44 = 164$ --- respectively).
The amplitudes $\tau(t)$, $f_{\log}(\tau_s)$, $f_0(\tau_s)$, $f_1(\tau_s)$, presented in Figs.\ \ref{fig: fit fm} and \ref{fig: fit sg}, were modeled as
combinations of rational functions, weighted with $1 / [1 + \exp(\pm a_t t)]$ (for the thermal amplitude) and $1 / [1 + \exp(\pm a \tau_s)]$ (for the main term and correction terms), as in:
\begin{equation}
  \tau(t) = \frac{\tau_p(t)}{1 + \exp(-a_t t)} + \frac{\tau_m(t)}{1 + \exp(+a_t t)},
\end{equation}
and
\begin{equation}
  f_i(\tau_s) = \frac{f_{i,p}(\tau_s)}{1 + \exp(-a \tau_s)} + \frac{f_{i,m}(\tau_s)}{1 + \exp(+a \tau_s)}.
\end{equation}
The $\tau_s$-dependence of the main term is monotonic (no ``dip'') in the ferromagnetic case. Observe the limited range of values $(0.0, 0.6)$ in the spin-glass case.

\begin{figure}[b]
\centerline{\includegraphics[width=0.98\columnwidth]{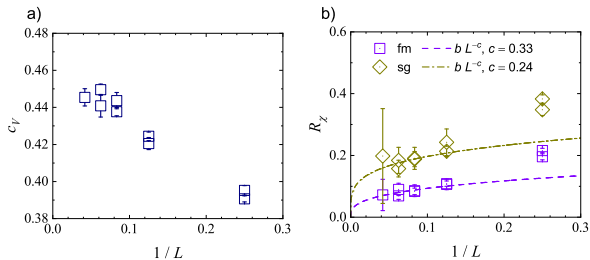}}
\caption{Size dependence of selected quantities at the criticality ($1 / T^{*} = 0.96$).
Left panel: specific heat; right panel: self averaging of susceptibilities.}
\label{fig: sa}
\end{figure}

In order to investigate the self-averaging behavior, $R_{\chi}(L, T^{*})$ was directly fitted (excluding $L = 4$) to the expression
$b L^{-c}$, yielding fits of adequate quality (judged by reduced $\chi^2$, cf.\ Fig.\ \ref{fig: sa}) and comfortably consistent
with self-averaging behavior in both the ferromagnetic and spin-glass sectors. The exponents for the fits, obtained from a hyperscaling relation, are
$c = 0.33$ and $0.24$ in the two sectors, respectively, where $c = -\alpha / \nu$ if self-averaging takes place.

However, the presence of a logarithmic correction in the spin-glass case suggests a special behavior, as
there is no implication $(R_{\chi} \to 0) \not\Rightarrow (V_4 - V_4' \to 0)$ in such case.
Indeed, with a logarithmically divergent $V_4' \to -\infty$, (\ref{eq: reltosa}) limits $V_4 - V_4'$ only to $o(\log(L))$.

\section{Irreducible components of the order parameter in 3D Heisenberg model}
\label{sec: decomp_q}

This section is devoted to the decomposition of the order parameter into irreducible components of the $O(3)$ symmetry group. If anisotropy is absent, the space of the spin degree of freedom is not related to the real space
in which the lattice is defined, and the cubic symmetry of the lattice does not affect the $O(3)$ one.
The order parameter $q_{\alpha\beta} = \sum_i s_{i\alpha} \otimes s_{i\beta}$, where $\alpha$ and $\beta$ stand for two independent replicas of the system (at the same temperature) and $i$ goes through the lattice sites,
is a rank-2 tensor under the transformations of the $O(3)$ group. It is therefore rather natural to consider $q$ as the total of the unitary ($q \propto \delta_{ab}$), antisymmetric ($q_{\alpha\beta} = -q_{\alpha\beta}$)
and traceless-symmetric ($q_{\alpha\beta} = q_{\beta\alpha}$, $q_{\alpha\alpha} = 0$) components. Since the components are mutually orthogonal, the norm $q^2$ is the total of their norms.

One observes further that
\begin{equation}
  \left< \left\| q_{\alpha\beta} \right\|^2 \right> = \sum_{i, j} \left< s_i \cdot s_j \right>_\alpha \left< s_i \cdot s_j \right>_\beta = \sum_{i, j} \left\| \left< s_i \cdot s_j \right> \right\|^2
\end{equation}
($\left\| \cdot \right\|^2$ stands for the square of the Euclidean norm of a tensor or a vector, or a scalar).
Dually, $q_d$ can be defined as
\begin{equation}
  q_{d,\alpha\beta} = \sum_i s_{i\alpha} \cdot s_{i\beta}
\end{equation}
and
\begin{eqnarray}
  \left< \left\| q_{d,\alpha\beta} \right\|^2 \right> & = & \sum_{i, j} \left( \left< s_i \otimes s_j \right>_\alpha \middle| \left< s_i \otimes s_j \right>_\beta \right) \nonumber \\ & = & \sum_{i, j} \left\| \left< s_i \otimes s_j \right> \right\|^2,
\end{eqnarray}
in which $\left( \cdot \middle| \cdot \right)$ is the scalar product (contraction) of same-valenced tensors (or vectors, or scalars), i.e. $\left( v \middle| v \right) = \left\| v \right\|^2$.
Even the vector cross product can be utilized, as in
\begin{equation}
  q_{c,\alpha\beta} = \sum_i s_{i\alpha} \times s_{i\beta}.
\end{equation}
Now, $s_i \otimes s_j$ can be decomposed into the isotropic part (proportional to the identity matrix and with zero angular momentum $\ell = 0$), the antisymmetric part with $\ell = 1$, and the traceless symmetric part ($\ell = 2)$. Naturally, the squared norm is the sum of the contributions,
\begin{eqnarray}
  \lefteqn{\left\| \left< s_i \otimes s_j \right> \right\|^2 = \left\| P_0 \left( \left< s_i \otimes s_j \right> \right) \right\|^2 + {}} \nonumber \\
   & & {} + \left\| P_1 \left( \left< s_i \otimes s_j \right> \right) \right\|^2 + \left\| P_2 \left( \left< s_i \otimes s_j \right> \right) \right\|^2,
\end{eqnarray}
where $P_\ell, \ell = 0, 1, 2$ stand for the corresponding projection operators.
In parallel, $q_d^2 = \sum_{\ell = 0}^2 q_\ell^{(2)}$, where we have introduced the notation:
\begin{eqnarray}
  q_0^{(2)} & = & \frac{1}{3} q^2, \\
  q_1^{(2)} & = & -\frac{1}{2} q^2 + \frac{1}{2} q_d^2 + \frac{1}{2} q_c^2, \\
  q_2^{(2)} & = & \frac{1}{6} q^2 + \frac{1}{2} q_d^2 - \frac{1}{2} q_c^2.
\end{eqnarray}
Although we have $\left< q_1^{(2)} \right> \approx \left< q_2^{(2)} \right> \approx 0$ in the whole range of temperatures near $T_c$, and thus $\left< q_d^2 \right> \approx \left< q_0^{(2)} \right> \propto \left< q^2 \right>$,
$\left< \left(q_1^{(2)}\right)^2 \right>$ and (appreciably) $\left< \left(q_2^{(2)}\right)^2 \right>$ do not vanish (thermal fluctuations are present).
This is why the Binder ratio is affected if $q^2$ is replaced with $q_d^2$.

Since the replicas of the system are independent, the symmetry group acts independently on each of them,
and the covariance tensor $C^{abcd}(i, j) = \mathop{\mathrm{cov}}(q^{ab}_{i,\alpha\beta}, q^{cd}_{j,\alpha\beta})$
is invariant with respect to the $O(3) \times O(3)$ symmetry group acting on the Cartesian indices $((a, c), (b, d))$.
We write:
\begin{eqnarray}
  \lefteqn{\mathop{\mathrm{cov}}(s^a_{i,\alpha}, s^b_{j,\alpha}) = \frac{1}{3} \left< s_i \cdot s_j \right>_{\alpha} \delta^{ab},} \\
  \lefteqn{\mathop{\mathrm{cov}}(q^{ab}_{i,\alpha\beta}, q^{cd}_{j,\alpha\beta}) = {}} \nonumber \\
  & & \quad  \mathop{\mathrm{cov}}(s^a_{i,\alpha}, s^c_{j,\alpha}) \mathop{\mathrm{cov}}(s^b_{i,\beta}, s^d_{j,\beta}), \label{eq: cov sg} \\
  \lefteqn{\left< s^a_{i,\alpha} s^b_{j,\alpha} s^c_{k,\alpha} s^d_{m,\alpha} \right> = {}} \nonumber \\
  & &  \quad \frac{1}{6} \biggl[ \left< (s_{i,\alpha} \cdot s_{j,\alpha}) (s_{k,\alpha} \cdot s_{m,\alpha}) \right> \delta^{ab} \delta^{cd} + {} \nonumber \\
  & &  \qquad \left< (s_{i,\alpha} \cdot s_{k,\alpha}) (s_{j,\alpha} \cdot s_{m,\alpha}) \right> \delta^{ac} \delta^{bd} + {} \nonumber \\
  & &  \qquad \left< (s_{i,\alpha} \cdot s_{m,\alpha}) (s_{j,\alpha} \cdot s_{k,\alpha}) \right> \delta^{ad} \delta^{bc} \biggr] + {} \nonumber \\
  & & \quad {} - \frac{1}{30} \biggl[ \left< (s_{i,\alpha} \cdot s_{j,\alpha}) (s_{k,\alpha} \cdot s_{m,\alpha}) \right> + {} \biggr. \nonumber \\
  & & \qquad\qquad \left< (s_{i,\alpha} \cdot s_{k,\alpha}) (s_{j,\alpha} \cdot s_{m,\alpha}) \right> + {} \nonumber \\
  & & \qquad\qquad \biggl. \left< (s_{i,\alpha} \cdot s_{m,\alpha}) (s_{j,\alpha} \cdot s_{k,\alpha}) \right> \biggr] \times {} \nonumber \\
  & & \qquad \qquad \qquad \left( \delta^{ab} \delta^{cd} + \delta^{ac} \delta^{bd} + \delta^{ad} \delta^{bc} \right). \label{eq: cov4}
\end{eqnarray}
Accordingly, in the simple model without an anisotropy, correlation lengths for the three components of $q_{\alpha\beta}$ coincide.
However, the ratio $\text{fm-}\xi_L / \text{sg-}\xi_L$ fails to be constant (and equal 2), indicating a departure from (\ref{eq: C x y z}). This is expected, as the distance dependence of the correlator is some combination of the exponential and power functions; in particular, it is power-law at the exact critical point. It follows by (\ref{eq: cov sg}) that the system with one replica (``the ferromagnetic sector'') and the system with two of them (``the spin-glass sector'') belong to different universality classes.

By a simple fourfold summation of (\ref{eq: cov4}) over the lattice sites, we obtain
\begin{widetext}
\begin{eqnarray}
  \lefteqn{\left< m^a_{\alpha} \, m^b_{\alpha} \, m^c_{\alpha} \, m^d_{\alpha} \right> -
    \left[ \left< m^a_{\alpha} \, m^b_{\alpha} \right> \left< m^c_{\alpha} m^d_{\alpha} \right> +
      \left< m^a_{\alpha} \, m^c_{\alpha} \right> \left< m^b_{\alpha} \, m^d_{\alpha} \right>
      \left< m^a_{\alpha} \, m^d_{\alpha} \right> \left< m^b_{\alpha} \, m^c_{\alpha} \right> \right] = {}} \nonumber \\
  & & -\frac{2}{5} \frac{5 \left( \left< m^2 \right> \right)^2 - 3 \left< \left( m^2 \right)^2 \right>}{2 \left( \left< m^2 \right> \right)^2}
    \left[ \left< m^a_{\alpha} \, m^b_{\alpha} \right> \left< m^c_{\alpha} m^d_{\alpha} \right> +
      \left< m^a_{\alpha} \, m^c_{\alpha} \right> \left< m^b_{\alpha} \, m^d_{\alpha} \right>
      \left< m^a_{\alpha} \, m^d_{\alpha} \right> \left< m^b_{\alpha} \, m^c_{\alpha} \right> \right],
\end{eqnarray}
where the l.h.s. is a joint cumulant, $\kappa_4\left( m^a_{\alpha}, m^b_{\alpha}, m^c_{\alpha}, m^d_{\alpha} \right)$,
and the r.h.s. is proportional to $\left( \delta^{ab} \delta^{cd} + \delta^{ac} \delta^{bd} + \delta^{ad} \delta^{bc} \right)$.
We write:
\begin{equation}
  \left[ \kappa_4\left( m^a_{\alpha}, m^b_{\alpha}, m^c_{\alpha}, m^d_{\alpha} \right) \right] =
    -\frac{2}{45} V_4'\{m\}  \left[ \left( \left< m^2 \right> \right)^2 \right]
    \left( \delta^{ab} \delta^{cd} + \delta^{ac} \delta^{bd} + \delta^{ad} \delta^{bc} \right).
\end{equation}
\end{widetext}

\section{Conclusions}

A disorder average prescription has been introduced into the definition of the Binder ratio, as needed for site-diluted systems of Heisenberg spins.
The effect of the lattice (here, fcc) on the calculated correlation lengths has been discussed and accounted for.
A required collapse of the $V_4'(\xi / L^y)$ dependencies onto one curve is evident for appropriately chosen values of $y$.
The departure of $y$ from the expected value ($y \ne 1$) may affect the determination of the transition temperature.

Now we come to the important question what is the self-averaging behavior of the model.
In fact, in the absence of self-averaging one would face the dilemma: which data to use, $V_4$ or $V_4'$?
Fortunately, $R_{\chi} \to 0$ and we have (\ref{eq: reltosa}).
Despite this convergence, the definition (\ref{eq: v4'}) appears as advantageous over (\ref{eq: v4}),
as the latter yields unwanted features which may interfere with the fitting procedure (presumably,
additional terms would need to be included in the fit if one chose $V_4$).

\section*{Acknowledgments}

The access to the computing facilities of the Interdisciplinary Center of Modeling at the University of Warsaw (Grant No. G68-12)
is kindly acknowledged. Mathematical optimization software packages (\texttt{ipopt} \cite{Wachter_2006}, hsl \cite{hsl})
were in use at intermediate stages. The author thanks Prof. Jerzy Wr\'obel for helpful discussion.

\appendix

\section{Algorithms}

\subsection{Heat bath}
\label{app: heat bath}

The local update algorithm \cite{Lee:2007} involves the evaluation of the expression $u$ ($0 \leq u \leq 1$)
\begin{equation}
  u(r) = - \frac{\log\left[ 1 + r (e^{-2 \beta h} - 1) \right]}{2 \beta h}
\end{equation}
[$0 \leq r \leq 1$ is a $U(0, 1)$-distributed random number,
$h = \left\| \mathbf{h} \right\|$ is the Euclidean norm of the magnetic field vector $\mathbf{h}$, and $\beta = \frac{1}{k_B T}$ the inverse temperature].
The danger of a floating-point exception or a catastrophic round-off error can be avoided if $u$ is rewritten as:
\begin{equation}
  u(r) = \frac{1}{\beta h} \mathop{\mathrm{arctanh}} \left[ \frac{-r (e^{-2 \beta h} - 1)}{2 + r (e^{-2 \beta h} - 1)} \right],
\end{equation}
applying the transformations $\beta h \to - \beta h$, $r \to 1 - r$, $u \to 1 - u$ if $\beta h < 0$, and
using the \texttt{expm1} library function to evaluate $e^x - 1$ accurately. If $\left| \beta h \right|$ is close to zero, $u \approx r$.

The heat bath algorithm requires also a rotation $R$ of the three-dimensional Cartesian coordinate system such that
$R \mathbf{e}_z = \mathbf{n} = n_x \mathbf{e}_x + n_y \mathbf{e}_y + n_z \mathbf{e}_z$. The appropriate rotation matrix reads
\begin{equation}
  R = \left( \begin{array}{ccc} 1+a& b& n_x\\ b& 1+c& n_y\\ -n_x& -n_y& n_z \end{array} \right),
\end{equation}
with
\begin{eqnarray}
  a & = & -\frac{n_x^2}{1 + n_z}, \\
  b & = & -\frac{n_x n_y}{1 + n_z}, \\
  c & = & -\frac{n_y^2}{1 + n_z},
\end{eqnarray}
and the transformations $\mathbf{n} \to -\mathbf{n}$, $h \to -h$ if $n_z < 0$.

The reformulated algorithm proceeds as follows:
\begin{enumerate}

\item generate of a uniformly distributed unit vector $\mathbf{\omega}$,
$\omega_z = 1 - 2 r$, $\omega_x = 2 \sqrt{r (1 - r)} \cos (2 \pi s)$, $\omega_y = 2 \sqrt{r (1 - r)} \sin (2 \pi s)$,
$0 \leq r, s \leq 1$ (alternatively, one normalizes to unity a vector from the three-dimensional multinormal distribution);

\item rotate the coordinate system, $\mathbf{\omega}' = R^{-1} \mathbf{\omega}$, according to the direction of the magnetic field $\mathbf{n} = \mathbf{h} / h$;

\item calculate $r' = (1 - \omega'_z) / 2$, and normalize the two-dimensional projection
$\omega_2 = (\omega_{2,x}, \omega_{2,y}) = (\omega'_x, \omega'_y) / \sqrt{{\omega'_x}^2 +  {\omega'_y}^2}$;

\item calculate $u = u(r')$ and $w = 2 \sqrt{u (1 - u)}$;

\item recompose a unit vector as $\mathbf{\omega}'' = (w \omega_{2,x}, w \omega_{2,y}, 1 - 2u)$;

\item apply the rotation $R$; the final spin direction is $R \mathbf{\omega}''$.

\end{enumerate}
Steps 1--4 guarantee continuity of the generated spin configuration with respect to $\beta$, $\mathbf{h}$, and the random data $(r, s)$.
This is desirable in view of the presence of round-off errors in floating-point calculations.

\subsection{Parallel tempering with Glauber probabilities}
\label{app: ptg}

In parallel tempering (PT), temperature replicas of the simulated systems are being interchanged with energy-dependent probabilities,
according to the Metropolis prescription: the move, from a configuration $C$ to the configuration $C'(i, i + i)$ with the replicas $(i, i + 1)$ one taking place of another,
is being accepted with the probability $p_M(\Delta \beta \, \Delta E)$. Taking into account the fact that the interchange of two replicas is an involution,
one can replace $p_M$ with Glauber's $p_G(\Delta \beta \, \Delta E)$. The detailed balance condition requires
\begin{equation}
  p_G(\Delta \beta \, \Delta E) / p_G(-\Delta \beta \, \Delta E) = e^{\Delta \beta \, \Delta E},
\end{equation}
which in view of the identity
\begin{equation}
  p_G(\Delta \beta \, \Delta E) + p_G(-\Delta \beta \, \Delta E) = 1
\end{equation}
amounts to
\begin{equation}
  p_G(\Delta \beta \, \Delta E) = 1 / (1 + e^{-\Delta \beta \, \Delta E}).
\end{equation}
The performance of the resulting algorithm depends on the regime in which the simulation is performed: if the number of thermal replicas $N_T$ is large,
the Metropolis algorithm (which favors interchange) performs better (the replicas travel faster on the temperature axis; ideally, the number of required PT steps is proportional $N_T$).
However, it is not unusual to optimize $N_T$ for the economy of the simulation, and then the \emph{variance of the replica random walk} $(1 - p) p$ is maximized by $p = 1 / 2 = p_G(0)$ ---
with the number of required Monte Carlo steps proportional to $N_T^2$, i.e.\ to the heat capacity. In contrast, $p_M(0) = 1$  yields a zero variance of the replica walk.

\subsection{Ratios of expected values}
\label{app: rat_avg}

Being given $n$ statistically independent samples $(x_i, y_i)$ of a two-variate statistical distribution on $(0, \infty) \times (-\infty, \infty)$,
we estimate the ratio of expected values $\left< y \right> / \left< x \right>$ in terms of $\mathtt{sigma}[k, m] = \sum_{i = 1}^n x_i^k y_i^m$ according to the expression (displayed below)
obtained by expanding
\begin{eqnarray}
  \lefteqn{1 / (x_0 + (x - x_0))} \nonumber \\
  & = & 1 / x_0 - (x - x_0) / x_0^2 + \ldots + O(x - x_0)^5, \nonumber
\end{eqnarray}
with $x_0 = \sum_{i = 1}^n x_i / n$. In limited testing the incurred bias shown as negligible,
unless $x$ and $y$ were significantly correlated (they usually are).

\begin{widetext}
\begin{verbatim}
(sigma[0,
     1] ((120 + n (-10 + n (35 + (-10 + n) n))) sigma[1, 0]^4 -
      n^2 (120 + (-10 + n) n) sigma[1, 0]^2 sigma[2, 0] -
      2 (-20 + n) n^3 sigma[1, 0] sigma[3, 0] +
      3 n^4 (sigma[2, 0]^2 - 2 sigma[4, 0])) +
   n ((-240 + n (20 + (-10 + n) n)) sigma[1, 0]^3 sigma[1, 1] +
      2 n (120 + (-10 + n) n) sigma[1, 0]^2 sigma[2, 1] -
      3 (-20 + n) n^2 sigma[1,
        0] (sigma[1, 1] sigma[2, 0] - 2 sigma[3, 1]) -
      4 n^3 (3 sigma[2, 0] sigma[2, 1] + 2 sigma[1, 1] sigma[3, 0] -
         6 sigma[4, 1])))/((-4 + n) (-3 + n) (-2 + n) (-1 + n) sigma[
    1, 0]^5)
\end{verbatim}
\end{widetext}

\bibliography{avg_br_4}

\end{document}